\AddToHook{package/hyperxmp/before}{\RequirePackage[bookmarksnumbered,unicode]{hyperref}}
\documentclass[9pt,sigconf,letterpaper]{acmart}
\usepackage{colortbl}
\colorlet{cpupath}{green!14}
\setkeys{acmart.cls}{balance=false}
\copyrightyear{2026}
\acmYear{2026}
\setcopyright{cc}
\setcctype{by-nc-nd}
\acmConference[CoNEXT-SW '26]{ACM CoNEXT 2026 Student Workshop}{December 07--10, 2026}{Utrecht, Netherlands}
\acmBooktitle{ACM CoNEXT 2026 Student Workshop (CoNEXT-SW '26), December 07--10, 2026, Utrecht, Netherlands}
\acmDOI{10.1145/3843179.3858061}
\acmISBN{979-8-4007-2976-8/2026/12}
\hypersetup{pdftitle={Workload-Aware Monitoring for AI-RAN}}
\title{Workload-Aware Monitoring for AI-RAN}
\author{Babak Azkaei}
\affiliation{%
  \institution{Eindhoven University of Technology}
  \city{Eindhoven}
  \country{Netherlands}}
\email{b.azkaei@tue.nl}

\author{Georgios Exarchakos}
\affiliation{%
  \institution{Eindhoven University of Technology}
  \city{Eindhoven}
  \country{Netherlands}}
\email{g.exarchakos@tue.nl}

\author{Kishor Chandra Joshi}
\affiliation{%
  \institution{Eindhoven University of Technology}
  \city{Eindhoven}
  \country{Netherlands}}
\email{k.c.joshi@tue.nl}
\input{arxiv_notice_conext.tex}

\begin{document}

\renewcommand{\shortauthors}{Azkaei et al.}

\begin{abstract}
AI-RAN platforms are expected to share CPU and GPU resources between radio workloads and AI services. Aggregate telemetry reports operating conditions, but timing changes may reflect different workloads, processors, or measurement boundaries. We propose workload-aware monitoring that associates radio-work descriptors with processing paths, timing boundaries, and operating context before comparing like with like. In 16 over-the-air downlink experiments, hybrid dispatch (CPU for transport blocks (TBs) below 4,096 bytes and GPU otherwise) has 29.66\% and 22.93\% lower mean host-callback durations than direct GPU dispatch, without and with a synthetic GPU co-runner, respectively. Grouping locates the separation in small TBs, while large-TB callback-duration ranges overlap. These observations identify a candidate for complementary CPU/GPU use.
\end{abstract}



\maketitle

\section{Introduction}
AI-RAN platforms run software radio processing on CPUs and GPUs~\cite{sionna-kit,ocudu-cuda}, and AI-and-RAN designs share these GPUs with AI services~\cite{yinyangran}. Choosing a processor for each radio task requires monitoring that explains where processing time goes. Telemetry usually exports aggregates, since per-task records are voluminous: one phone in our testbed produces about 1{,}400
downlink transport blocks (TBs) per second. But aggregates can mislead. We showed that letting the CPU process small blocks (hybrid approach) lowers mean host-callback durations by about a quarter. Did the GPU get faster, did the
workload change, or did work move to the CPU? The average alone cannot
tell. Benchmarks also show that accelerator gains shrink for small tasks, as an example in LDPC decoding~\cite{decodex}. Systems such as CloudRIC~\cite{cloudric} coordinate heterogeneous RAN resources; our focus is interpreting workload-conditioned observations.

GPU traces~\cite{cupti} and resource managers~\cite{yinyangran} provide
building blocks. What is missing is a RAN-level layer that keeps, for each
unit of work, \emph{what was processed, where it ran, what was timed and
under which conditions}, revealing what benchmarks cannot: in a running
base station, the scheduler sets task sizes and the task mix shifts. We
propose such \emph{workload-aware monitoring} and illustrate it on
over-the-air physical downlink shared channel (PDSCH) processing.

\section{Workload-Aware Monitoring}\label{sec:framework}
\begin{figure}[t]
\centering
\includegraphics[width=\columnwidth]{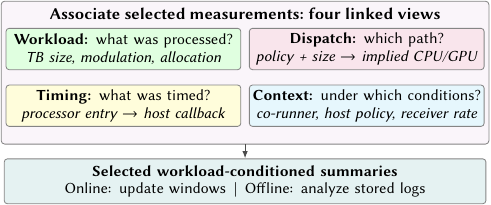}
\caption{Workload-aware monitoring (italics: Our PDSCH use case).}
\Description{Four views associate selected measurements with workload,
configured dispatch, timing and context. PDSCH examples are TB size,
modulation and allocation; policy plus size implying CPU or GPU;
processor entry to host callback; and co-runner, host policy and receiver
rate. Selected workload-conditioned summaries may be updated online or
computed offline. The PDSCH case uses online collection and offline analysis.}
\label{fig:framework}
\end{figure}

The unit of analysis is a \emph{work item}, such as a PDSCH encoding
request. For each item, the framework keeps four linked views
(\emph{workload}, \emph{dispatch}, \emph{timing} and \emph{context}), each
answering one question (Figure~\ref{fig:framework}). They draw on native
logs and counters, run records and, optionally, user-defined telemetry: a
Janus codelet~\cite{janus} or CUPTI records~\cite{cupti} could add the
processor that actually ran an item and when its GPU output was ready. Native logs already pair TB descriptors with callback duration here.

Two rules turn the views into monitoring output. \emph{Associate before
aggregating}: link records through shared work identifiers or fields
logged together, not nearby timestamps alone. \emph{Aggregate like with like}: per context, report the share and timing of each workload class on each processor (e.g., a duration histogram labeled by TB size and processor), not one overall mean. An overall change can then be traced to a different workload mix, a different processor, or a timing change within comparable work.

Online monitoring would update bounded summaries as records arrive; offline analysis applies the associations to stored logs. \newline
Only workload relevant key performance indicators (KPIs) receive chosen groupings: Here we condition callback duration on TB class and policy. This case uses online collection and offline analysis.

\section{Illustrative PDSCH Case}\label{sec:case}
\begin{figure}[t]
\centering
\includegraphics[width=\columnwidth]{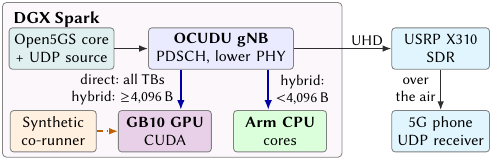}
\caption{Testbed. Black: downlink traffic. Blue: where the compared PDSCH
stage runs under each policy. Dash-dot: the optional co-runner's GPU work.}
\Description{A DGX Spark hosts the Open5GS core with a UDP source, the OCUDU gNB, a synthetic co-runner, the GB10 GPU and the Arm CPU. Downlink traffic flows from the core to the gNB, through a USRP X310 radio and over the air to a 5G phone. Under direct dispatch the PDSCH stage sends all TBs to the GPU; under hybrid dispatch TBs below 4,096 bytes go to the CPU and larger TBs to the GPU. The co-runner also submits work to the GPU.}
\label{fig:testbed}
\end{figure}

\textbf{Testbed and policies.} An NVIDIA DGX Spark (Arm CPU, GB10 GPU) runs the Open5GS core and a CUDA-enabled OCUDU base station (gNB)~\cite{ocudu-cuda} (Figure~\ref{fig:testbed}). A USRP X310 radio serves one phone over a 20-MHz single-antenna (1$\times$1) link in 5G NR band n77; a UDP source sends 30~Mbps for 20~s, and the gNB's scheduler sets the TB sizes. At the compared PDSCH codeword-processing stage, \emph{direct} runs every TB on the GPU, while \emph{hybrid} runs small TBs (below 4{,}096 bytes) on the CPU and large ones on the GPU; the rest of the workflow is similar. To emulate AI-and-RAN GPU sharing, a synthetic co-runner is configured for nominal 1-ms GPU work followed by 9-ms host sleep.

\textbf{Mapping to the views.} Live logs pair TB size, modulation and
allocation with the steady-clock processor-entry-to-callback-entry
interval (0.1-$\mu$s rounding). This \emph{host-callback duration} excludes
pre-entry queueing and logger submission, not all internal waiting.
GPU callbacks can precede output readiness. Policy and size imply,
rather than trace, execution; run settings and receiver reports supply context.

\textbf{Population.} We retain 223{,}979 single-codeword callbacks for
the target phone from 16 selected successes, four per setting.
Their later producer-log timestamps select [5,15)~s after each traffic-control
request, not exact callback-entry times.

\begin{table}[t]
\caption{Callback duration: minimum--maximum of the per-run averages. Small: TB $<4{,}096$~B. Large share: range of large/all retained-record fractions per run. Shading: CPU-selected small cells; hybrid all-TB cells mix paths.}
\label{tab:ota}
\centering
\small
\begin{tabular}{@{}l@{\hspace{5pt}}rrrr@{}}
\toprule
 & \multicolumn{3}{c}{Mean callback duration ($\mu$s)} & Large \\
\cmidrule(lr){2-4}
Policy & All TBs & Small TBs & Large TBs & share \\
\midrule
\multicolumn{5}{@{}l}{\textit{Without co-runner}} \\
Direct & 49.4--52.5 & 48.3--52.5 & $52.3$--$56.6$ & 0.0--26.5\% \\
Hybrid & 34.3--38.2 & \cellcolor{cpupath}28.5--38.0 & 53.5--56.3 & 0.9--23.2\% \\
\addlinespace[3pt]
\multicolumn{5}{@{}l}{\textit{With GPU co-runner}} \\
Direct & 48.8--51.9 & 47.8--51.8 & $53.5$--$61.1$ & 0.0--32.3\% \\
Hybrid & 37.0--39.8 & \cellcolor{cpupath}29.2--39.5 & 51.3--54.6 & 1.5--31.7\% \\
\bottomrule
\end{tabular}
\end{table}

\section{Results}\label{sec:results}
\textbf{Hybrid's observed advantage.} Table~\ref{tab:ota} groups each
run-wide policy's workload by size; the columns are not additional modes.
Comparing equal-weight averages of run means, not range midpoints,
hybrid's all-TB callback durations are 29.66\%/22.93\% lower
without/with the co-runner. Small-TB means are 36.09\%/30.29\% lower
where hybrid selects the CPU, while retaining CUDA for larger blocks.
This locates an observed callback time advantage. 

\textbf{Workload composition.} The ``Large share'' column shows that
the proportion of large TBs varies across runs. Comparing results
within the same size category does not make the workloads identical. Each direct-policy large-TB range uses three runs because the fourth
contains no large-TB records. Overlapping callback-duration ranges do not establish equal GPU performance. The logs do not identify whether processing delay, buffering or link adaptation caused the different TB-size proportions.

\textbf{Scope.} Client-returned receiver reports give near 30~Mbps and zero reported UDP loss over approximately 20-s intervals. Enabled-logger/export counters report zero loss.

\section{Conclusion}
In our OTA PDSCH case, workload-aware analysis locates lower small-TB host-callback durations under hybrid dispatch, where the configured path uses the CPU. These observations motivate complementary CPU/GPU use. Next steps are an online implementation, device-side completion timing (e.g., via CUPTI), a cutoff sweep, and more stages such as uplink shared-channel (PUSCH) decoding.
More broadly, such monitoring could drive online CPU/GPU dispatch that adapts to load and co-located AI work.

\begin{acks}
This work is partially supported by the European Union's Horizon Europe Marie Sk\l{}odowska-Curie Action SCION under Grant Agreement No. 101072375 and the NL National Growth Fund 6G flagship project Future Network Services.
\end{acks}

\bibliographystyle{ACM-Reference-Format}
\bibliography{references}
\end{document}